\documentclass[12pt]{article}
\usepackage{epsfig}
\begin{document}
\title{\bf Multiplicities of Periodic Orbit Lengths for Non-Arithmetic Models}
\author{{\it E. Bogomolny,  and C. Schmit}\\
 Laboratoire de Physique Th\'eorique et Mod\`eles Statistiques\\
 Universit\'e de Paris XI, B\^at. 100\\
 91405 Orsay Cedex, France}

\maketitle

\begin{abstract}
Multiplicities of periodic orbit lengths for non-arithmetic Hecke
triangle groups are discussed.  It is demonstrated both numerically and
analytically that at least for certain groups the mean
multiplicity of periodic orbits with exactly the same length increases
exponentially with the length.  The main ingredient used 
is the construction of joint distribution of periodic orbits when group
matrices are transformed by field isomorphisms. The method can be
generalized to other groups for which traces of group matrices are integers
of an algebraic field of finite degree.
\end{abstract}

\section{Introduction}
For chaotic systems the density of classical periodic orbits
with a given length increases exponentially. In particular, for all constant 
negative curvature surfaces generated by discrete groups one has the universal
asymptotics (see e.g. \cite{Hejhal})
\begin{equation}
\rho_{\mbox{\scriptsize total}}(l)\stackrel{l\to\infty}{\longrightarrow}
\frac{e^{l}}{l}\;.
\label{Huber}
\end{equation}
Much less is known about multiplicities of periodic orbits with exactly the
same length. Usually it is assumed that the mean length multiplicity
of periodic orbits for generic systems  depends only on exact symmetries and
for models without geometrical symmetries
the mean multiplicity $\bar{g}$ equals $2$ or $1$ for systems
respectively with or without time-reversal invariance. 

Physically it means that, in general,  there exists no reason that two 
different periodic orbits would have the same length except for time-reversal
invariant systems where almost all trajectories can be traversed  in two 
opposite directions which implies that $\bar{g}=2$.
In semiclassical approach to spectral statistics of chaotic systems the
distinction between these two classes of models is reflected in different
behaviour of the two-point correlation form factor at the origin which agrees
with the predictions of the random matrix theory \cite{Berry}, \cite{Berry2}.

For the free motion on constant negative curvature surfaces generated by
discrete groups the situation is different. In such hyperbolic models
periodic orbits are in one-to-one correspondence with conjugacy classes of
group matrices and the length of a periodic orbit, $l_p$, is directly related
with the trace of a matrix $M$ representing each class  (see e.g. \cite{Hejhal})
\begin{equation}
|\mbox{Tr} M|=\left \{ 
  \begin{array}{cl} 
    2\cosh l_p/2\;,&\mbox{ if }\det M=1\\
    2\sinh l_p/2\;,&\mbox{ if }\det M=-1 
  \end{array}\right .  \;.
\label{length}
\end{equation}
Hence, any relations between traces of group matrices imply
connections between periodic orbit lengths.

The extreme case corresponds to the so-called arithmetic groups (see e.g.
\cite{BBGS} and references therein). For such groups 
traces of group matrices can take only quite restricted set of values and 
the number of possible traces less than a given value  is asymptotically 
\cite{BBGS}
\begin{equation}
N(|\mbox{Tr} M|<X)\stackrel{X\to\infty}{\longrightarrow}
CX
\end{equation}
with a system dependent constant $C$.  Because
$X\stackrel{l\to\infty}{\longrightarrow}e^{l/2}$ but not all possible values
of traces really appear for group matrices, the number of periodic orbits with different
lengths when $l_p\to \infty$ has the following upper bound \cite{BBGS}
\begin{equation}
N_{\mbox{\scriptsize diff.}}(l_p<l)\leq Ce^{l/2}\;.
\end{equation}
Define the mean multiplicity of periodic orbit length as the ratio of
the density of all periodic orbit to the density of periodic orbits with
different lengths
\begin{equation}
\bar{g}(l)=\frac{\rho_{\mbox{\scriptsize total}}(l)}
{\rho_{\mbox{\scriptsize diff.}}(l)}
\end{equation}
where 
$\rho_{\mbox{\scriptsize diff.}}(l)=d N_{\mbox{\scriptsize diff.}}(l_p<l)/dl$.

From the above formulas one proves \cite{BBGS} that for arithmetic groups the 
mean multiplicity is exponentially large and has the following estimate from 
below
\begin{equation}
\bar{g}(l)\geq \frac{2e^{l/2}}{Cl}\;.
\label{arithmetic}
\end{equation}
In classical mechanics such large multiplicities play a minor role but their
interference 
changes drastically quantum mechanics of arithmetic groups. In particular,
spectral statistics of arithmetic systems is close to the Poisson statistics
typical for integrable systems and not to the random matrix statistics
conjectured for chaotic models \cite{BBGS}, \cite{Steiner}, \cite{BLS}. 

Arithmetic systems are very exceptional but according to the
Horowitz--Randol theorem \cite{Horowitz}, \cite{Randol} for all hyperbolic
models generated by discrete groups multiplicities are unbounded.
Nevertheless, multiplicities covered by this theorem are quite rare and
a priori  assumption would be that for non-arithmetic hyperbolic models the mean 
multiplicity equals 2 as for generic time-reversal invariant systems.

Numerical calculations performed in \cite{BBGS} indicated that it is not
always the case. In that paper certain non-arithmetic Hecke triangles were
considered and it was observed that mean multiplicity of periodic orbits
with length $l$ seems to increase exponentially
\begin{equation}
\bar{g}(l)\sim e^{\lambda l}
\end{equation}
with an exponent $\lambda<1/2$. 

The purpose of this paper is two-fold. First, we perform numerical
calculations  of periodic orbits for much larger lengths that in \cite{BBGS} 
and, second, we develop a method
which gives a lower bound of multiplicities, thus in certain cases
confirming analytically exponential growth of multiplicities.

The plan of the paper is the following. In Section \ref{code} we discuss
general properties of Hecke triangle group matrices.
In Section~\ref{Numerical} results of numerical calculations of
periodic orbit length multiplicities for a few Hecke triangles are presented. As
traces of Hecke triangle group matrices are integers of an algebraic field,
each group matrix defines not one but a few different lengths corresponding
to different isomorphisms of the basis field. In Section~\ref{isomorphisms}
the construction of the joint distribution for periodic orbits with all
transformed lengths fixed is discussed. In Section~\ref{DifferentLengths} it is
demonstrated how the knowledge of this joint distribution permits to
calculate the lower bound of the periodic orbit length multiplicity. In
Section~\ref{simplest} the computations are performed for the simplest case
of Hecke groups with $n=5\;,\;8\;,\;10\;,\;12$ which are characterized by
the existence of only one non-trivial isomorphism. In Appendix A it is 
proved that for Hecke triangle groups  all transformed lengths are smaller
than the true length. This inequality is sufficient to ensure 
that for Hecke groups with only one non-trivial isomorphism
length multiplicities increases exponentially. Our results agree
well with direct numerical computations of periodic orbit multiplicity for
these groups. In Section~\ref{generalCase} other Hecke groups are shortly 
considered. It appears that in all investigated cases except groups with one
non-trivial isomorphism  length multiplicities increase so slowly
that the direct check is practically impossible. In Section~\ref{statistics} 
we briefly discuss the influence of periodic orbit length
multiplicities on the spectral statistics for corresponding systems. 
In Section~\ref{summary} a
summary of the results is given. In Appendix B a saddle point
method of calculation of the joint distribution of periodic orbit lengths is
discussed.

\section{Hecke triangles}\label{code}

Hecke triangles are hyperbolic triangles with angles $0,\pi/2,\pi/n$ with
integer $n\geq 3$. All of them are fundamental regions of discrete groups
$G_n$ generated by reflections across its sides. Let us denote the reflection
across the side connecting angles $0$ and $\pi/2$ by $A$, the one across the
side connecting angles $0$ and $\pi/n$ by $B$ and the last one by $C$. From
geometrical considerations these transformations obey the defining relations
\begin{equation}
A^2=B^2=C^2=1\;,\;(AC)^2=(BC)^n=-1\;.
\label{grammar}
\end{equation}
The explicit form of $A$, $B$, and $C$ can be chosen as follows
\begin{equation}
A=\left (\begin{array}{cc}-1&0\\0&1\end{array}\right )\;,\;
B=\left (\begin{array}{cc}-1&2\cos \pi/n\\0&1\end{array}\right )\;,\;
C=\left (\begin{array}{cc}0&1\\1&0\end{array}\right )\;.
\label{ABC}
\end{equation}
Arbitrary matrix from the Hecke triangle group $G_n$ is a word of these letters.
Due to (\ref{grammar}) these symbols have a complicated grammar.
For our purposes it is convenient to introduce new symbols 
\begin{eqnarray}
\alpha_1^m&=&C(AB)^m\;,\nonumber\\
\alpha_2^m&=&CB(AB)^m\;,\nonumber\\
\alpha_3^m&=&CBC(AB)^m\;,\\
\ldots &&\ldots \nonumber\\
\alpha_{n-2}^m&=&\underbrace{CBCB\ldots }_{n-2\;\mbox{\scriptsize symbols}}
(AB)^m \nonumber
\end{eqnarray}
where $m=1,2,\ldots $ are positive integers. 

Explicitly up to unessential overall sign
\begin{equation}
\alpha_{2k+1}^m=
\left (\begin{array}{cc}-a_{k}&\alpha m a_{k}+a_{k-1} \\
      -a_{k+1}&\alpha m a_{k+1}+a_{k} \end{array}\right ),\;
\alpha_{2k}^m=
\left (\begin{array}{cc}-a_{k-1}&\alpha m a_{k-1}+a_{k} \\
-a_{k}&\alpha m a_{k}+a_{k+1} \end{array}\right )
\label{explicitCode}
\end{equation}
where from now we denote  
\begin{equation}
\alpha=2\cos \pi/n
\label{alpha}
\end{equation}
and $a_k\equiv a_k(\alpha)$ are the Chebyshev polynomials of the second kind
of $\alpha$ 
\begin{equation}
a_k=\frac{\sin (k\pi/n)}{\sin (\pi/n)}\;.
\label{ak}
\end{equation}
Using the defining relations (\ref{grammar})
one can proves \cite{BBGS} that conjugacy classes in $G_n$ (and, consequently, 
periodic orbits in Hecke triangles) can be constructed as free words in these 
new symbols with the only restriction that cyclic permutations correspond to 
the same orbit. 

Due to specific form of generators (\ref{ABC}) matrix elements of the Hecke
triangular group matrices are polynomials with integer coefficients of the
variable $\alpha\equiv 2\cos \pi/n$,  thus forming naturally a subfield of 
the cyclotomic field of degree $2n$. 

The constant $\alpha$ defined in (\ref{alpha}) obeys a polynomial 
equation $P_N(\alpha)=0$ with integer coefficients
\begin{equation}
P_N(x)=\prod_{\begin{array}{c}k=\mbox{\small odd }\\ (k,n)=1 
\end{array}}
(x-\alpha_k(n) )=x^N+\ldots 
\end{equation}
where 
\begin{equation}
\alpha_k(n)=2\cos(\pi k/n)
\label{alphak}
\end{equation}
and the product is taken over all odd integer $k$ coprime with $n$. 
The total number of such integers and, consequently, the degree of the
defining equation is  
\begin{equation}
N=\frac{1}{2}\varphi(2n)
\end{equation}
where $\varphi(p)$ is the  Euler $\varphi$-function which 
counts the number of integers less than $p$ and coprime with $p$.

In Table~\ref{equations} the explicit forms of the defining polynomials 
for low values of $n$  are presented.   In the last column of this table we
give for  later use the discriminant of these polynomials  defined as the
square of the product of all roots
\begin{equation}
\Delta_n= \prod_{k<m<N}\left [\alpha_k(n)-\alpha_m(n)\right ]^2
\label{discriminant}
\end{equation}
where $\alpha_k(n)$ is given by (\ref{alphak}), and the product is taken 
over all odd integers $k<m$ both coprime with $n$.
For even $n$ and odd $k$ $\alpha_{n-k}(n)=-\alpha_k(n)$, and 
\begin{equation}
\Delta_{n}=2^N\Delta_n^{(e)}\Delta_n^{(o)}
\end{equation}
where $\Delta_n^{(e,o)}$ are the discriminants of even and odd  
powers of $\alpha$
\begin{eqnarray}
\Delta_n^{(e)}&=&\prod_{k<m\leq N/2}\left [\alpha_k^2(n)-\alpha_m^2(n)\right ]^2
\label{de}\;,\\
\Delta_n^{(o)}&=&\prod_{k<m\leq N/2}\left [\alpha_k(n)\alpha_m(n)
(\alpha_k^2(n)-\alpha_m^2(n))\right ]^2\;.
\label{do}
\end{eqnarray}

\begin{table}
\caption{\small Irreducible monic polynomials defining the field
  $\mbox{ l\hspace{-.47em}Q}(2\cos\pi/n)$ for small $n$. $N$ is the degree
  of the polynomial,  $\Delta_n$ is its discriminant. For $n$ even
  the discriminant is given as the product of 3 terms. The second and third
  factors represent discriminants for even and odd functions (see (\ref{de})
  and (\ref{do})).}
\begin{center}
\begin{tabular}{|c||c|c|c|}
\hline\noalign{\smallskip}
$n$ & $N$ & $P_N(x)$&$\Delta$\\
\noalign{\smallskip}\hline\noalign{\smallskip}
$5$ & $2$ & $x^2-x-1$ & $5$\\
$7$ & $3$ & $x^3-x^2-2x+1$ & $7^2$\\
$8$ & $4$ & $x^4-4x^2+2$ & $2^4\cdot 2^3\cdot 2^4$\\
$9$ & $3$ & $x^3-3x -1$ & $3^4$\\
$10$ & $4$ & $x^4-5x^2+5$ & $2^4\cdot 5\cdot 5^2$\\
$11$ & $5$ & $x^5-x^4-4x^3+3x^2+3x-1$& $11^4$\\
$12$ & $4$ & $x^4-4x^2+1$ & $2^4\cdot 12 \cdot 12$ \\
$13$ & $6$ & $x^6-x^5-5x^4+4x^3+6x^2-3x-1$ & $13^4$\\
$14$ & $6$ & $x^6-7x^4+14x^2-7$ & $2^6\cdot 7^2 \cdot 7^3$\\
$15$ & $4$ & $x^4+x^3-4x^2-4x+1$ & $3^2 5^3$\\\hline
\end{tabular}
\end{center}
\label{equations}
\end{table}

Therefore all matrix elements and, in particular, traces of Hecke group matrices 
have the following form
\begin{equation}
\mbox{ Tr} M=\sum_{k=0}^{N-1}n_k\alpha^k
\label{integer}
\end{equation}
with integer coefficients $n_k$.

As matrix elements of Hecke groups are algebraic integers of the totally real 
field  $\mbox{ l\hspace{-.47em}Q}(2\cos\pi/n)$ it is natural to consider in
parallel all isomorphisms  of this field defined by  the following
substitutions
\begin{equation}
\varphi_k:\;\;\;\alpha \longrightarrow \alpha_k=2\cos\frac{\pi k}{n}
\label{iso}
\end{equation}
for all odd integers $k<n$ coprime with $n$.

In general, the number of such isomorphisms equals the degree of the
defining polynomial but in our case $\alpha$ and $-\alpha$ both correspond to
the same group. Hence, when this transformation belongs to the group of
isomorphisms (which is the case for even $n$), it does not change periodic 
orbit lengths. Consequently, the dimension of the group of isomorphisms of 
periodic orbit lengths, $q$, is 
\begin{equation}
q=\left \{ \begin{array}{cc}N,&\mbox{ for  odd $n$}\\
                 \frac{1}{2}N,&\mbox{ for even $n$}\end{array}\right . \;.
\end{equation}			    
In particular, the following four cases corresponds to the simplest case of 
the groups of isomorphisms of the order 2 (cf. Table~\ref{equations})
\begin{equation}
n=5\;,\;8\;,\;10\;,\;12\;. 
\label{two}
\end{equation}
It appears that multiplicities of periodic orbit lengths depend strongly on
the number of isomorphisms so we consider first the case (\ref{two}).

\section{Numerical Calculations}\label{Numerical}

\begin{figure}
\begin{center}
\epsfig{file=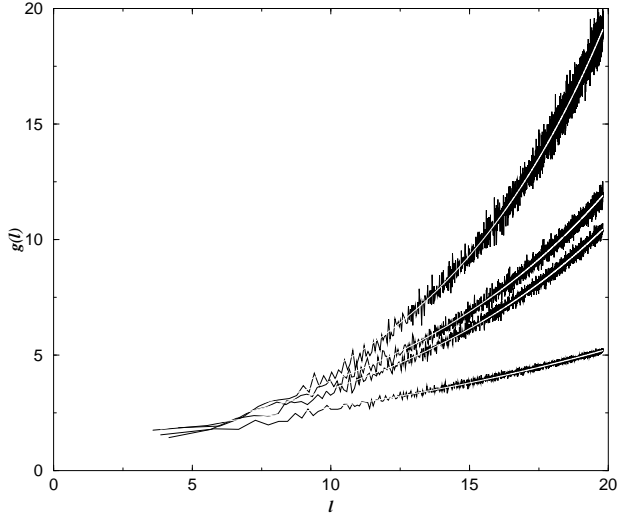, width=7cm, angle=-90}
\end{center}
\caption{\small Mean multiplicities of periodic orbit lengths for
Hecke triangles  $(0,\pi/2,\pi/n)$ with (from top to bottom)
$n=12$, $n=5$, $n=8$, and $n=10$ for $l<20$. 
White lines are  numerical fits  (\ref{5})--(\ref{12}).}
\label{g}
\end{figure}

At Fig.~\ref{g} we present the numerically computed multiplicity for the
Hecke triangles  $(0,\pi/2,\pi/n)$ with $n=5,8,10,12$ for lengths $l<20$. 
White lines indicate a two-parameter fit to these data in the form 
$\bar{g}(l)\approx a_n e^{b_n l}$
\begin{eqnarray}
n&=&5\ \;:\;\;\;\;\bar{g}(l)\approx 1.235 e^{0.114 l}\;,\label{5}\\
n&=&8\ \;:\;\;\;\;\bar{g}(l)\approx 1.095 e^{0.114 l}\;,\label{8}\\
n&=&10\;:\;\;\;\;\bar{g}(l)\approx 1.143 e^{0.065 l}\;,\label{10}\\
n&=&12\;:\;\;\;\;\bar{g}(l)\approx 0.986 e^{0.150 l}\;.\label{12}
\end{eqnarray}

\begin{figure}
\begin{center}
\epsfig{file=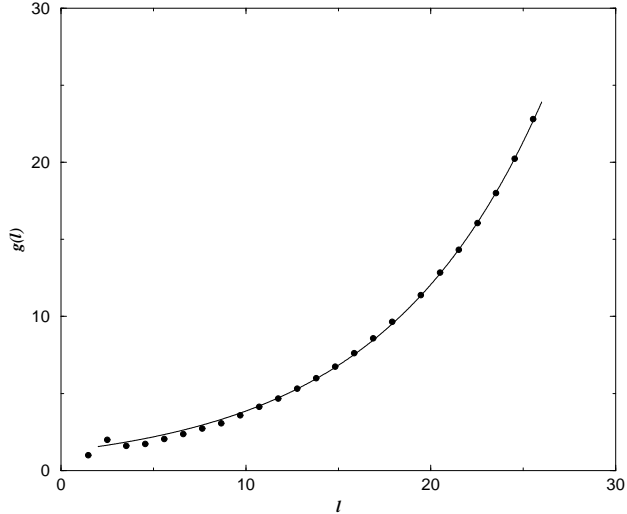, width=7cm, angle=-90}
\end{center}
\caption{\small Mean multiplicities of periodic orbit lengths for the
Hecke triangle  with angles $(0,\pi/2,\pi/5)$ for $l<25$. Solid line is the
fit (\ref{5}). }
\label{gll}
\end{figure}
Expressions (\ref{5})--(\ref{12}) fit numerical data in the given interval of 
lengths pretty well. But they are purely best least-square numerical 
fits and no attempts were made to determine the accuracy of coefficients. 
In Section~\ref{simplest} it is 
demonstrated that our approach suggests  different formulas for these
quantities (see (\ref{corr})) which, nevertheless, are practically 
indistinguishable from the above simple expressions in the considered interval
of lengths (cf. Fig.~\ref{fitt}).  

For larger lengths the exponential proliferation of
periodic orbits makes it difficult to compute and store in the memory all
periodic orbits. Nevertheless the determination of periodic orbits in a
reasonably short interval of lengths is still feasible. 
At Fig.~\ref{gll} we present the result of the numerical computation of
the length multiplicity for the Hecke triangle with $n=5$ till $l=25$. Each
small circle at this figure for $l>20$ corresponds to one million of periodic 
orbits. The solid line is the fit (\ref{5}) obtained from data at small $l$. 
It is clearly seen that accuracy of the fit does not change noticeably with 
the increasing of periodic orbit lengths.

\section{Length distribution for different\\ isomorphisms}\label{isomorphisms}

For the Hecke triangle groups (and for certain other groups as well) 
traces of group  matrices are integers of an algebraic field of finite degree. 
Therefore each group matrix $M$ gives rise not only to one usual length 
(\ref{length}) but to $q$ different lengths corresponded to $q$ different
isomorphisms of the basis field applied to a matrix $M$. Asymptotically 
\begin{equation}
l_k=2\ln |\mbox{Tr }\varphi_k(M)|\;.
\end{equation}
In this definition $l_1$ corresponding to the identity transformation is the
true length of a periodic orbit and all other $l_k$ with $k\geq 2$ are
additional quantities which we call transformed lengths. 

For arithmetic systems (see e.g. \cite{BBGS}) transformed traces are
restricted
\begin{equation}
|\mbox{Tr }\varphi_k(M)|\leq 2
\end{equation}
for all $k\geq 2$ which leads to the very large length multiplicities 
lengths for such groups (\ref{arithmetic}).

The main ingredient of our approach to the problem of length multiplicity
for non-arithmetic groups is the determination of the joint density of periodic
orbits in intervals $l_k,l_k+dl_k$ for all $k\geq 1$. 
For clarity we first consider  groups with only one non-trivial
isomorphisms (\ref{two}) where each hyperbolic group matrix permits to 
define two lengths, $l_1$ and $l_2$. 

Let $R(l_1,l_2)dl_1dl_2$ be the number of periodic orbits with the first length
in the interval $l_1,l_1+dl_1$ and the second (transformed) length in the
interval $l_2,l_2+dl_2$. Taking into account (\ref{Huber}) one concludes that 
\begin{equation}
R(l_1,l_2)\approx \frac{e^{l_1}}{l_1}P(l_1,l_2)
\label{joint}
\end{equation}
where $P(l_1,l_2)$ has the meaning of the probability density of periodic
orbits with lengths $l_1$ and $l_2$ normalized such that 
\begin{equation}
\int_{-\infty}^{\infty}P(l_1,l_2)dl_2=1\;.
\label{normalization}
\end{equation}
No general arguments determining the form of $P(l_1,l_2)$ are known to the
authors. As $l_1$ is only one fixed quantity with the
dimension of the length, from physical considerations it is quite natural 
to assume that for large $l_1$ and $l_2$  this function has the following
scaling form (see also Appendix B for another argument)
\begin{equation}
P(l_1,l_2)=A(l_1)\exp \left [l_1f(l_2/l_1)\right ]
\label{scaling}
\end{equation}
with a certain (smooth) scaling function $f(u)$ where $u$ is the ratio of
two lengths.

\begin{figure}
\begin{center}
\epsfig{file=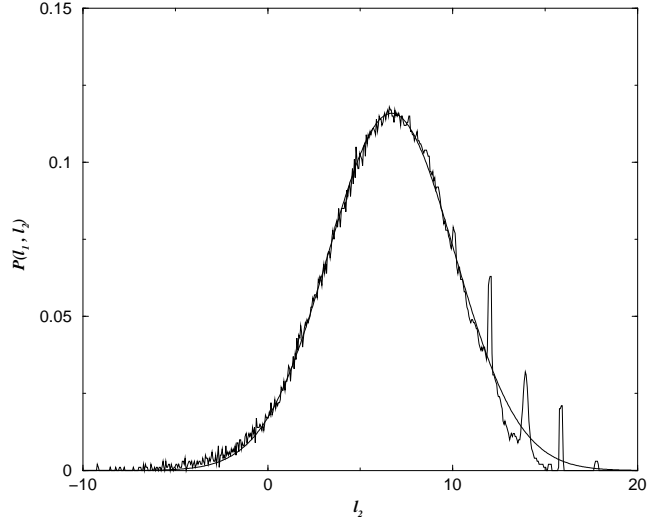, width=7cm, angle=-90}
\end{center}
\caption{\small Probability density of transformed length with $l_1=19.8$ for the
  Hecke triangle with $n=5$. Thick line represents the Gaussian fit 
  (\ref{Gaussian}),   (\ref{aaa}). }
\label{pll}
\end{figure}

\begin{figure}
\begin{center}
\epsfig{file=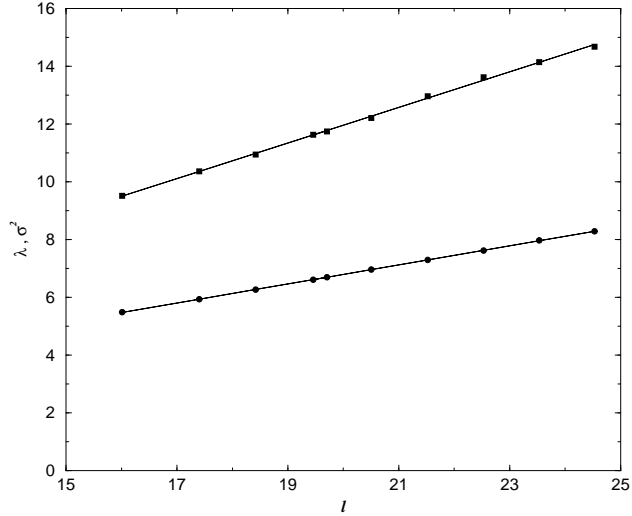, width=7cm, angle=-90}
\end{center}
\caption{\small Gaussian fit parameters (\ref{Gaussian}) versus 
  the length of periodic orbits for  the Hecke triangle with $n=5$. 
  Lower line: $\lambda$. Upper
  line: $\sigma^2$. Solid lines are the linear fits (\ref{bbb}) to these data.}
\label{mean}
\end{figure}

When $l_1\to \infty$ the prefactor $A(l_1)$ can be determined 
in the saddle point approximation from the normalization condition 
(\ref{normalization}) 
\begin{equation}
A(l_1)=\frac{1}{\sqrt{2\pi \sigma^2 l_1}}
\label{prefactor}
\end{equation}
where $\sigma^2=1/|f^{\prime \prime}(u_c)|$. Here $u_c$ is  
the point of the maximum of $f(u)$:
$f^\prime(u_c)=0$, and $f^{\prime \prime}(u_c)$ is the second
derivative of the function $f(u)$ at this point. From
(\ref{normalization}) it follows that the value of $f(u)$ at the point of
the maximum is zero
\begin{equation}
f(u_c)=0\;.   
\label{maximum}
\end{equation}
At Fig.~\ref{pll} we present numerically computed function $P(l_1,l_2)$ for
the Hecke triangle with $n=5$ for $10^6$ orbits near  $l_1\approx 19.8$
(which corresponds to $|\mbox{Tr }M|=20000$) 
together with the Gaussian fit to the data in the form
\begin{equation}
P(l_1,l_2)=a_0\exp \left (-\frac{(l_2-\lambda)^2}{2\sigma^2}\right )\;.
\label{Gaussian}
\end{equation}
The least square fit gives the following values of the parameters 
\begin{equation}
a_0=.116\;,\;\lambda=6.69\;,\;\sigma^2=11.74\;.
\label{aaa}
\end{equation}
At Fig.~\ref{mean} the best fit values of $\lambda(l_1)$ and $\sigma^2(l_1)$ 
are given for different values of $l_1$. The data are linear on $l_1$ 
and can be approximated by  the following straight lines
\begin{equation}
\lambda=.330 l_1+.187\;,\;\;\sigma^2=.616l_1-.337
\label{bbb}
\end{equation}
which support the scaling ansatz (\ref{scaling}). 

The peaks at  Fig.~\ref{pll} correspond to words in the code
(\ref{explicitCode}) with a small number of letters $\alpha_k^m$ but with
big  values of $m$'s. Ignoring all elements except the ones multiplied by
the largest possible numbers of  $m$'s one  can approximate 
the periodic orbit length as follows 
\begin{equation}
l\approx 2\ln (\alpha^p\  m_1\ldots m_p\ a_{k_{1}}\ldots a_{k_{p}})\;.
\end{equation}
Hence, in this approximation the difference between transformed lengths 
and  the true length is a finite constant
\begin{equation}
l_k-l_1\approx 2\ln (\varphi_k(\alpha^p\ a_{k_{1}}\ldots a_{k_{p}}))-
2\ln (\alpha^p\ a_{k_{1}}\ldots a_{k_{p}})\;.
\label{peaks}
\end{equation}
To compute the joint distribution of transformed lengths one considers
periodic orbits with the true length confined in a small interval. The above
expression means that orbits corresponding to small numbers of initial symbols
have transformed lengths at finite distances from $l_1$ and, consequently,
they  produce peaks at these distances. The quality of such approximation quickly
deteriorates  with increasing of $p$ due to the omitting lower powers of $m$'s
and in real calculations only a few peaks with small $p$ are visible.

For the Hecke triangle with $n=5$ $a_k$ defined in (\ref{ak}) are either $1$ 
or $\alpha$, and  all differences between two lengths are
\begin{equation}
l_2-l_1\approx 2 m\ln \left (\frac{\sqrt{5}-1}{\sqrt{5}+1}\right )\approx
-1.92\  m
\end{equation}
with integer $m$ which agree well with the  positions of the peaks at
Fig.~\ref{pll}. 

\begin{figure}
\begin{center}
\epsfig{file=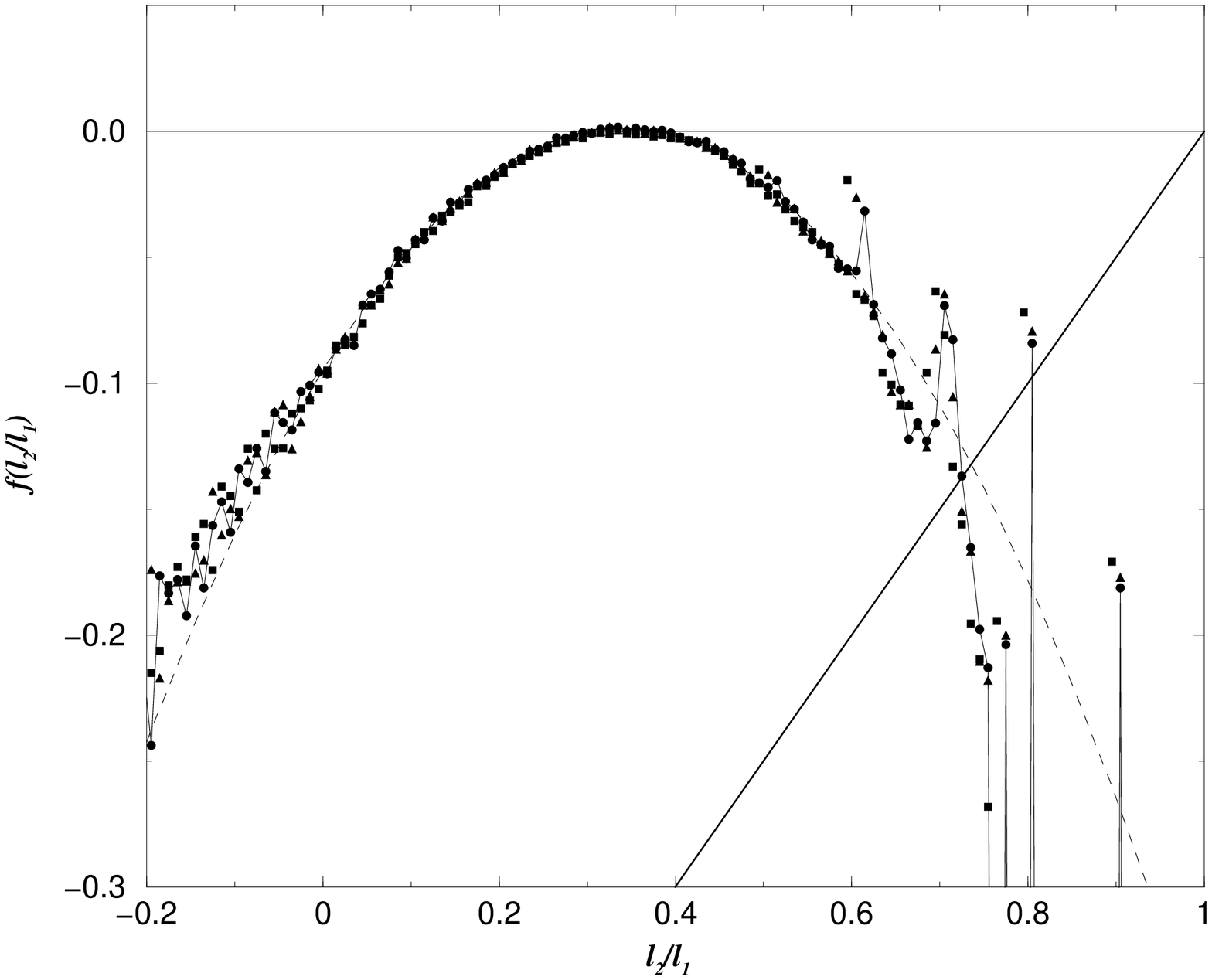, width=7cm, angle=-90}
\end{center}
\caption{ \small Scaling function $f(l_2/l_1)$ for the Hecke triangle with
  $n=5$. Circles, triangles, and squares represent data for $10^6$ orbits near
  respectively $l_1\approx 19.8\;,\;l_1\approx 19.47\;,\;l_1\approx 19.02$. Solid line connects
  points with $l_1\approx 19.8$. Dashed line is the
  parabolic fit (\ref{parabolic5}) to the data with $l_1\approx 19.8$. 
  Thick solid line is the  straight line $y=(x-1)/2$. }
\label{f5}
\end{figure}

\begin{figure}
\begin{center}
\epsfig{file=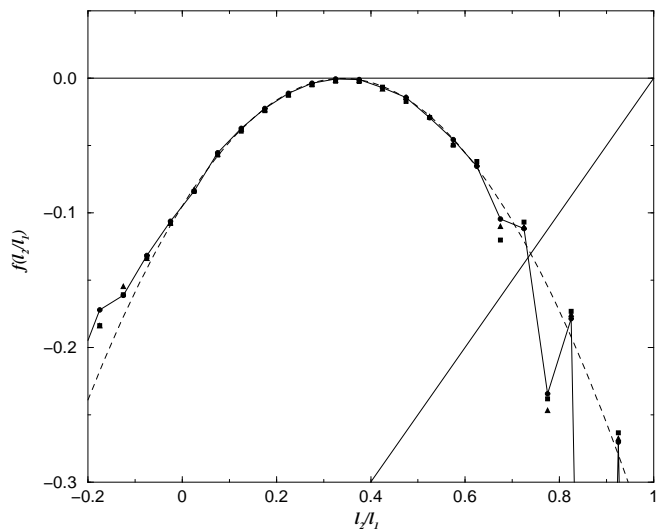, width=7cm, angle=-90}
\end{center}
\caption{\small The same as at Fig.~\ref{f5} but for the Hecke triangle 
  with $n=8$.}
\label{f8}
\end{figure}

\begin{figure}
\begin{center}
\epsfig{file=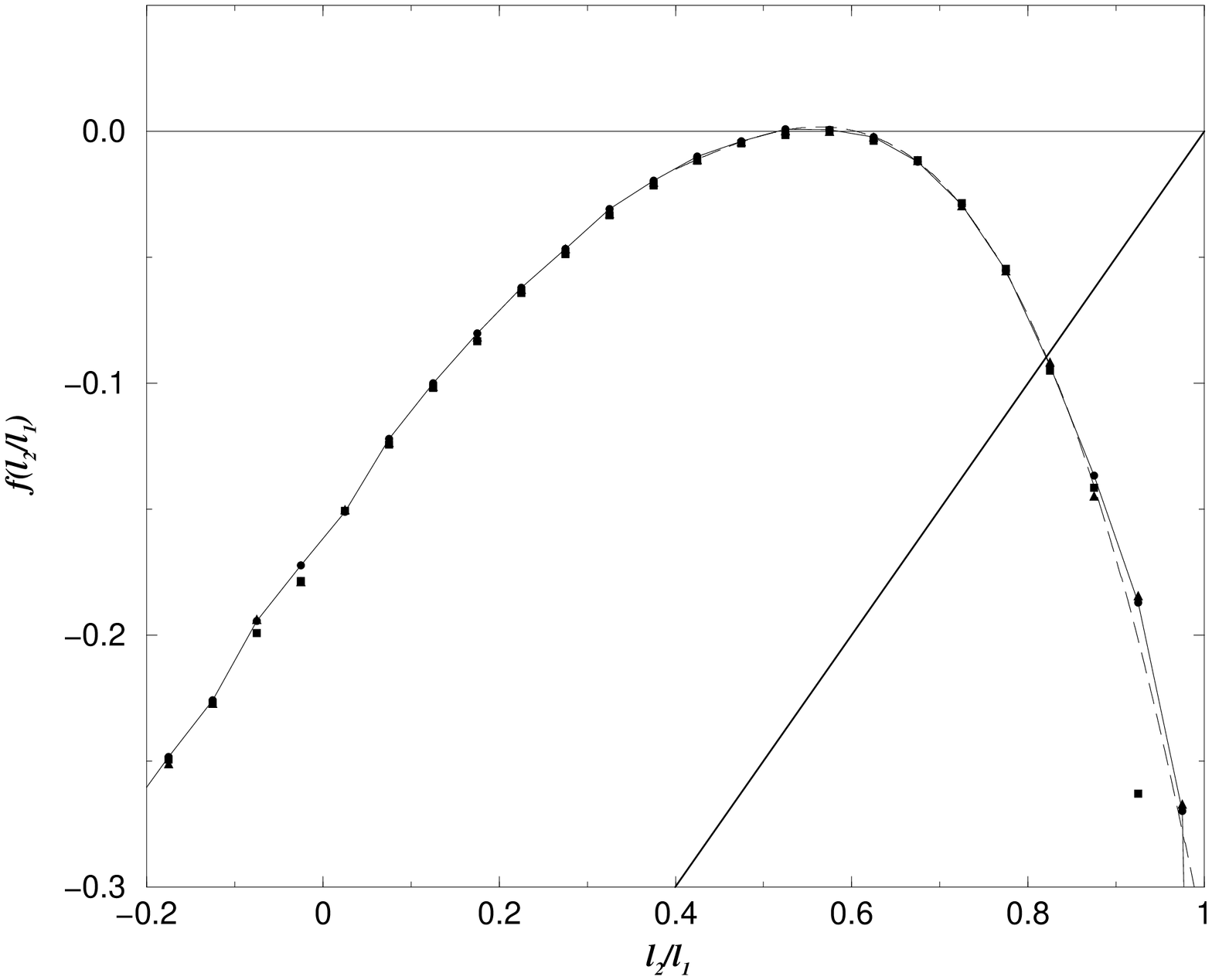, width=7cm, angle=-90}
\end{center}
\caption{\small The same as at Fig.~\ref{f5} but for the Hecke triangle with
  $n=10$. Dashed line is the cubic fit (\ref{cubic10}) to the data with 
  $l_1\approx 19.8$  in the interval $[.4,1]$.}
\label{f10}
\end{figure}

\begin{figure}
\begin{center}
\epsfig{file=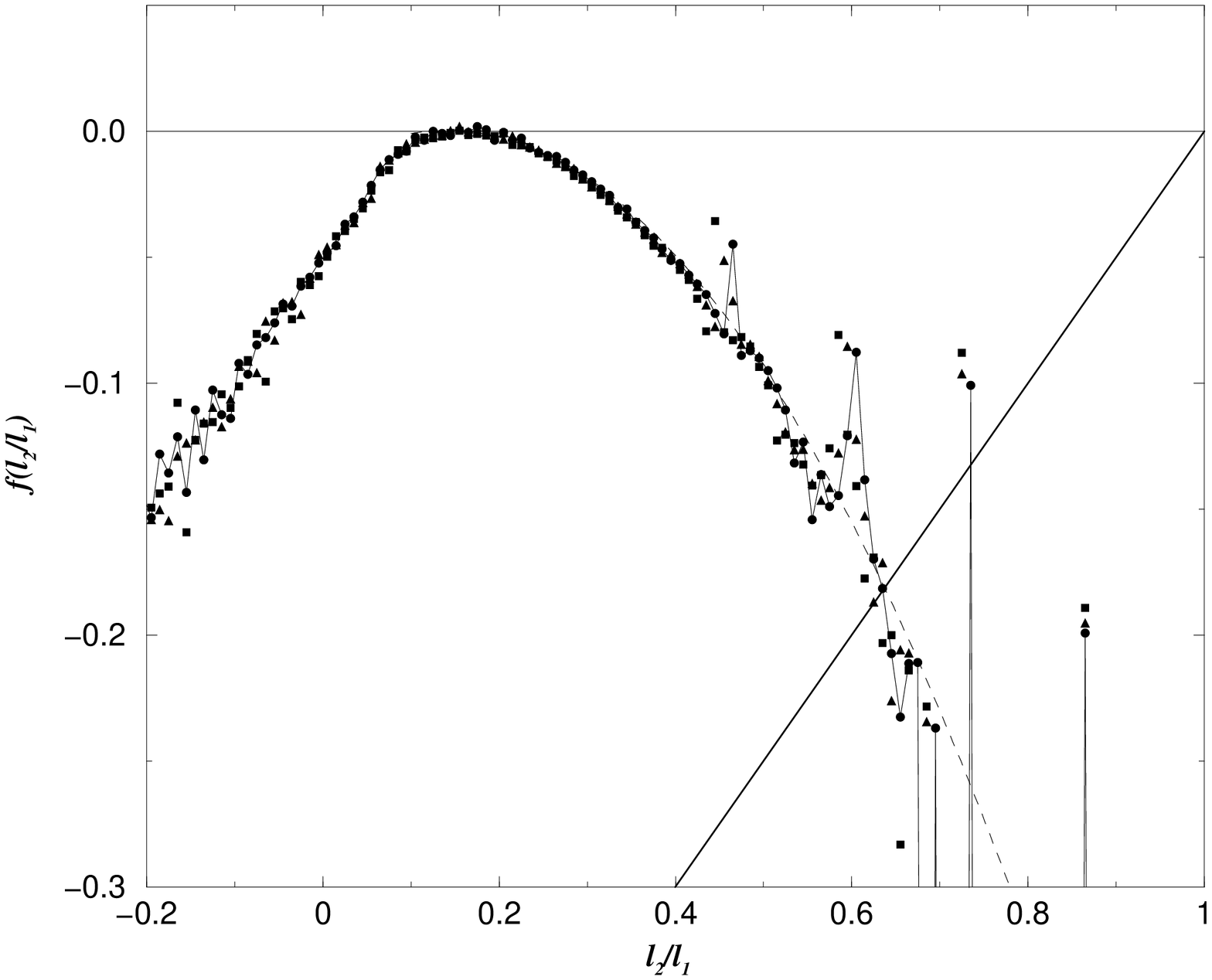, width=7cm, angle=-90}
\end{center}
\caption{\small The same as at Fig.~\ref{f5} but for the Hecke triangle with
  $n=12$. Dashed line is the parabolic fit (\ref{parabolic12}) to the data
  with $l_1\approx 19.8$ in the interval $[.1,1]$.}
\label{f12}
\end{figure}

At Figs.~\ref{f5}-\ref{f12} we plot numerically computed scaling functions
$f(u)$ for the Hecke triangles with $n=5$, $n=8$, $n=10$, and $n=12$ for
different intervals of periodic orbit lengths.    
The curves for different lengths seem to be superimposed thus supporting the
scaling ansatz (\ref{scaling}). Irregular points at $l_2/l_1\approx 1$
correspond to the above mentioned peaks (\ref{peaks}) related with words
with small number of symbols and are irrelevant at large $l_1$.

The scaling functions $f(l_2/l_1)$ for the Hecke triangles with $n=5$ and
$n=8$ are close to each other and can be reasonably well  described by the
following parabolic fit 
\begin{equation}
f(x)\approx -.094+.56x-.83x^2
\label{parabolic5}
\end{equation}
indicated by dashed lines at Figs.~\ref{f5} and \ref{f8}.

The scaling functions for the Hecke triangles with $n=10$ and $12$
have more complicated form. At Fig.~\ref{f10} the dashed line indicates the
cubic fit to the data in the interval $[.4,1]$
\begin{equation}
f(x)\approx .028-.66x+2.08x^2-1.77x^3\;.
\label{cubic10}
\end{equation}
At Fig.~\ref{f12} the dashed line shows the parabolic fit in the interval
$[.1,1]$
\begin{equation}
f(x)\approx -.014+.21x-.73x^2\;.
\label{parabolic12}
\end{equation}
For Hecke triangle groups with $n$ different from (\ref{two}) there exist more than 
one non-trivial
isomorphisms and, consequently, the joint distribution of all lengths have
the form similar to (\ref{joint}) but with larger number of transformed lengths   
\begin{equation}
R(l_1,l_2,\ldots ,l_q)\approx \frac{e^{l_1}}{l_1}P(l_1,l_2,\ldots ,l_q)\;.
\label{multijoint}
\end{equation}
The analog of the scaling ansatz (\ref{scaling}) in this case is
\begin{equation}
P(l_1,l_2,\ldots ,l_q)=A(l_1)\exp \left [ l_1
f\left (\frac{l_2}{l_1},\ldots ,\frac{l_q}{l_1}\right )
\right ]\;
\label{scalingGeneral}
\end{equation}
with a certain function $f(x_2,\ldots ,x_q)\equiv f(\vec{x})$ depended only 
on ratios $x_k=l_k/l_1$ and in the saddle point approximation 
\begin{equation}
A(l_1)=\sqrt{\frac{|\det \partial^2 f/\partial x_i\partial x_j|}
  {(2\pi l_1)^{q-1}}}
\end{equation}
where the derivatives are taken at the point of the maximum of $f(\vec{x})$.

\section{Number of periodic orbits with different
  lengths}\label{DifferentLengths} 

The importance of the knowledge of the joint distribution of periodic orbit
lengths for all possible isomorphisms is related with the fact that two
periodic orbits for Hecke triangle groups have exactly the same length iff
all their transformed lengths are the same.    

Let us consider the simplest Hecke group with $n=5$.
In this case the traces of group matrices, $t_1$ and $t_2=\varphi_2(t_1)$ 
can be written as 
\begin{equation}
t_1=n_0+n_1\lambda_1\;,\;\;t_2=n_0+n_1\lambda_2
\end{equation}
where $n_0,n_1$ are integers, $\lambda_1=2\cos(\pi/5)$ is an element of our 
basis field and $\lambda_2=2\cos(3\pi/5)$ is the transformed value of 
$\lambda_1$.

These equations determine  the transformation from variables $t_1,t_2$ to
variables $n_0,n_1$ and 
\begin{equation}
dt_1dt_2=Jdn_0dn_1
\end{equation}
where the Jacobian of this transformation is the square root of the
discriminant (\ref{discriminant}) of the defining equation
\begin{equation}
J=|\lambda_2-\lambda_1|=\sqrt{\Delta_5}=\sqrt{5}\;.
\end{equation}
As $t_i=e^{l_i/2}$ the precedent equation gives
\begin{equation}
dn_0dn_1=C_5e^{l_1/2+l_2/2}dl_1dl_2
\end{equation}
with $C_5=1/(4\sqrt{\Delta_5})$.

Because $n_0$ and $n_1$ are integers this equation means that in a volume
$dl_1dl_2$ there are at most $[Ce^{l_1/2+l_2/2}]$ possible values of
$n_0,n_1$ ($[x]$ is the integer part of x). This relation 
signifies that the density of the maximal number of periodic orbits with 
{\bf different} lengths obeys  asymptotically the inequality
\begin{equation}
\rho_{\mbox{\scriptsize diff.}}(l_1,l_2)\leq C_5e^{l_1/2+l_2/2}\;.
\label{different}
\end{equation}
We stress that such arguments can give, in principle, the estimate from above
because not all values of $n_0$ and $n_1$ are possible for the Hecke group
$G_n$, otherwise one obtains the Hilbert modular groups which are discrete
groups only in higher dimensional complex planes.

For other Hecke triangle groups with one non-trivial isomorphism (\ref{two}) 
(i.e.  for $n=8,10,12$) the defining
equation is of degree 4 but traces of group matrices contain either even or
odd powers of $\alpha$ 
\begin{equation}
t_1=n_0+n_2\alpha^2\;,\;\mbox{ or }\;t_1=n_1\alpha+n_3\alpha^3  
\end{equation}
and the result is similar to (\ref{different}) 
\begin{equation}
\rho_{\mbox{\scriptsize diff.}}(l_1,l_2)\leq C_ne^{l_1/2+l_2/2}
\label{different_general}
\end{equation}
but with
\begin{equation}
C_n=\frac{1}{4\sqrt{\Delta_n^{(e)}}}+\frac{1}{4\sqrt{\Delta_n^{(o)}}}
\end{equation}
where $\Delta_n^{(e,o)}$ are discriminants (\ref{de}) and (\ref{do}).

For general Hecke group with $q$ isomorphisms
\begin{equation}
\rho_{\mbox{\scriptsize diff.}}(l_1,l_2,\ldots ,l_q)\leq C_ne^{l_1/2+l_2/2+\ldots +l_q/2}\;
\label{diffrho}
\end{equation}
where for odd $n$
\begin{equation}
C_n=\frac{1}{2^q\sqrt{\Delta_n}}\;,
\end{equation}
and for even $n$
\begin{equation}
C_n=\frac{1}{2^q\sqrt{\Delta_n^{(e)}}}+\frac{1}{2^q\sqrt{\Delta_n^{(o)}}}\;.
\end{equation}
Eq.~(\ref{multijoint}) means that in a volume $dl_1\ldots dl_q$ there is 
\begin{equation}
\rho_{\mbox{\scriptsize tot.}}(\vec{l}\ )=\frac{e^{l_1}}{l_1}A(l_1)\exp\left (
l_1f(l_2/l_1,\ldots ,l_q/l_1)\right )
\end{equation}
periodic orbits  with all transformed lengths fixed. On the other hand in the
same volume the maximum
number of periodic orbits with different lengths is restricted by the 
inequality (\ref{diffrho})
\begin{equation}
\rho_{\mbox{\scriptsize diff.}}(\vec{l}\ )\leq C_ne^{l_1/2+l_2/2+\ldots +l_q/2}\;.
\end{equation}

Consequently, the maximum number of periodic orbits with different lengths is  
\begin{equation}
\rho_{\mbox{\scriptsize diff. lengths}}^{(\mbox{\scriptsize maximum})}(l_1)=
\int dl_2\ldots dl_q \left \{\begin{array}{cc}
\rho_{\mbox{\scriptsize diff.}}(\vec{l}\ )&
\mbox{ if  }\;\rho_{\mbox{\scriptsize diff.}}(\vec{l}\ )\leq 
\rho_{\mbox{\scriptsize tot.}}(\vec{l}\ )\\
\rho_{\mbox{\scriptsize tot.}}(\vec{l}\ )& 
\mbox{ if  }\;\rho_{\mbox{\scriptsize diff.}}(\vec{l}\ )\geq 
\rho_{\mbox{\scriptsize tot.}}(\vec{l}\ )
\end{array}\right .\;.
\label{rhomaximum}
\end{equation}
As  both densities increase exponentially with $l_1$  the dominant 
contribution to this integral is given by vicinities of boundary points where 
\begin{equation}
\rho_{\mbox{\scriptsize diff.}}(\vec{l}\ )=
\rho_{\mbox{\scriptsize tot.}}(\vec{l}\ )\;.  
\label{intersection}
\end{equation}
In the leading order of $l_1$ these points are determined from the equality
of the exponential factors of these functions
\begin{equation}
l_1+l_1f(l_2/l_1,\ldots ,l_q/l_1)=\frac{1}{2}(l_1+\ldots +l_q)\;.
\end{equation}
Denoting $l_k/l_1$ by $x_k$ one gets the equation independent of $l_1$
\begin{equation}
f(x_2,\ldots ,x_q)=\frac{1}{2}(x_2+\ldots +x_q-1)\;.
\label{mainq}
\end{equation}

\subsection{Groups with one non-trivial isomorphism}\label{simplest}

In the simplest case of groups (\ref{two}) where only one transformed length 
exists Eq.~(\ref{mainq}) is reduced to the equation of one variable 
$x\equiv x_2$
\begin{equation}
f(x)=\frac{1}{2}(x-1)\;.
\label{equation}
\end{equation}
In Appendix A it is proved that for the Hecke groups the transformed
lengths corresponding to all non-trivial isomorphisms are smaller than the
true length 
\begin{equation}
l_k< l_1\;.
\label{inequality}
\end{equation}
Consequently, $f(x)$ is situated at the  left from the line $x=1$ and as $f(u_c)=0$
$f(x)$ is negative when $x<1$. As $u_c<1$ Eq.~(\ref{equation})  for groups
with one non-trivial isomorphism always has a solution $x<1$.
In Table~\ref{parameters} we present approximate values
of this intersection point, $x_n$, for different values of $n$ found from 
Figs.~\ref{f5}--\ref{f12}.
\begin{table}
\caption{\small Parameters for Hecke triangles with $n=5\;,\;8\;,\;10\;,\;12$.
  The second column is the curvature in the point of the maximum.
  The third column gives the value of $C_n$ in (\ref{different_general}).
  The fourth and the fifth columns are the ordinate of the intersection point
  and  the modulus of the slope of $f(x)$ at this point. The nest three 
  columns  are  parameters in (\ref{fittheoric}). The last column gives the
  numerically computed prefactor in (\ref{corr}).}
\begin{center}
\begin{tabular}{|c||c|c||c|c||c|c|c||c|}
\hline\noalign{\smallskip}
$n$  & $\sigma_n^2$ & $C_n$ & $x_n$&$k_n$&$\lambda_n$&$\nu_n$&$G_n$& $K_n$\\
\noalign{\smallskip}\hline\noalign{\smallskip}
$5$  & $.6 $ & $.11 $ & $.74$ & $.66 $ & $.13 $ & $.35 $ & $ 1.32 $ & $ 2.34 $\\
$8$  & $.6 $ & $.15 $ & $.74$ & $.66 $ & $.13 $ & $.35 $ & $ 1.11 $ & $ 2.22 $\\
$10$ & $.55$ & $.16 $ & $.82$ & $.82 $ & $.09 $ & $.43 $ & $ 1.22 $ & $ 2.46 $\\
$12$ & $.68$ & $.14 $ & $.64$ & $.73 $ & $.18 $ & $.39 $ & $ 1.25 $ & $ 1.77 $\\\hline
\end{tabular}
\end{center}
\label{parameters}
\end{table}
As claimed in all these cases the solution exists and $x_n<1$.

\begin{figure}
\begin{center}
\epsfig{file=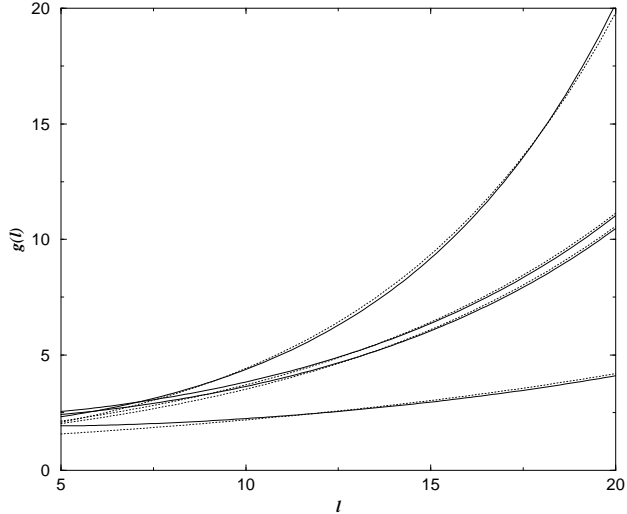, width=7cm, angle=-90}
\end{center}
\caption{\small Comparison of numerically computed fits (\ref{5})--(\ref{12}) 
  for  length multiplicities for Hecke
  triangles (dotted lines) with formulas (\ref{corr}) with fitted prefactor
  (solid lines). From top to bottom: $n=12$, $n=5$, $n=8$, $n=10$.}
\label{fitt}
\end{figure}

In the next order one can write
\begin{equation}
l_2=x_nl_1+\varepsilon_n\;.
\end{equation}
Expanding Eq.~(\ref{intersection}) to the first order of $\varepsilon$ one
gets
\begin{equation}
C_n\sqrt{2\pi \sigma_n^2 l_1^3}=\exp (-\varepsilon_n (k_n+.5)+
{\cal  O}(\varepsilon_n^2/l_1))
\end{equation}
where $k_n=|f^{\prime}(x_n)|$ is the modulus of the derivatives at the point
of the intersection. Therefore
\begin{equation}
\varepsilon_n=-\frac{1}{k_n+.5}
\ln \left (C_n\sqrt{2\pi \sigma_n^2}l_1^{3/2}\right )\;.  
\end{equation}
Together these formulas demonstrate that for the Hecke triangles (\ref{two}) 
at the intersection point 
\begin{equation}
\rho_{\mbox{\scriptsize diff.}}(l_1,l_2)=
\rho_{\mbox{\scriptsize tot.}}(l_1,l_2)\approx D_ne^{l_1(1+x_n)/2}l_1^{-3\beta_n/2}
\end{equation}
where
\begin{equation}
\beta_n=\frac{1}{2k_n+1}\;,\;\;
D_n=\frac{C_n^{1-\beta_n}}{(2\pi\sigma_n)^{\beta_n/2}}\;.
\end{equation}
The integration in (\ref{rhomaximum}) in the limit of large $l_1$ can be
performed by parts and finally
\begin{equation}
\rho_{\mbox{\scriptsize diff. lengths}}^{(\mbox{\scriptsize maximum})}(l_1)=
\frac{2D_n}{1-\beta_n}\frac{e^{l_1(1+x_n)/2}}{l_1^{3\beta_n/2}}\;.
\label{diff_lengths}
\end{equation}
The mean multiplicity of periodic orbit lengths is the ratio of the total
density of periodic orbits to the density of orbits with different lengths.
Hence
\begin{equation}
\bar{g}(l)\geq G_n\frac{e^{\lambda_n l}}{l^{\nu_n}}
\label{correct}
\end{equation}
where
\begin{equation}
\lambda_n=\frac{1-x_n}{2}\;,\;\;\nu_n=1-\frac{3}{2}\beta_n\;,\;\;
G_n=\frac{1-\beta_n}{2D_n}\;.
\label{fittheoric}
\end{equation}
At Table~\ref{parameters} we present approximate values of these parameters
computed from Figs.~\ref{f5}--\ref{f12}. At Fig.~\ref{fitt} we compare data
of length multiplicities for the Hecke triangles
(\ref{5})--(\ref{12}) with the formula of the form (\ref{correct}) 
\begin{equation}
\bar{g}(l)=K_n\frac{e^{\lambda_n l}}{l^{\nu_n}}
\label{corr}
\end{equation}
with the computed values of $\lambda_n$ and $\nu_n$ from
Table~\ref{parameters} but with a prefactor $K_n$ calculated from the best
fit to the data (see the last column of Table~\ref{parameters}).
The `theoretical' curves (\ref{corr}) are practically indistinguishable from
the best fits (\ref{5})--(\ref{12}). Note that fitted prefactors is always
bigger than $G_n$, just confirming that estimates (\ref{correct}) and
(\ref{fittheoric}) give only lower bounds. Though in principle not all
integers are allowed in (\ref{integer}), these results seem to indicate that in
the mean the ratio of the density of allowed integers to all integers for 
the groups (\ref{two}) is finite.   

\subsection{General case}\label{generalCase}

For general case of $q>2$ isomorphisms the arguments, in principle, remain 
the same. One has to perform the following three steps:
\begin{itemize}
\item to check that required solutions of Eq.~(\ref{mainq}) do exist,
\item to find on $(q-2)$-dimensional manifold of these solutions a point
  with the maximum of the sum $x_2+\ldots +x_q$,
\item to compute the integral (\ref{rhomaximum}) in a small vicinity of the
  point of the maximum.
\end{itemize}

For groups with only one non-trivial isomorphism the inequality (\ref{inequality}) was
sufficient to ensure the existence of a solution of Eq.~(\ref{mainq}). For
other groups it is not the case and one has to rely mostly on numerical
calculations. For example, the necessary condition of the existence of
solutions of Eq.~(\ref{mainq}) is that at the point $u_2,\ldots ,u_q$ of the
maximum of the  scaling function $f(x_2,\ldots ,x_q)$ the sum
$u_2+\ldots +u_q$ is less than $1$. 

\begin{figure}
\begin{center}
\epsfig{file=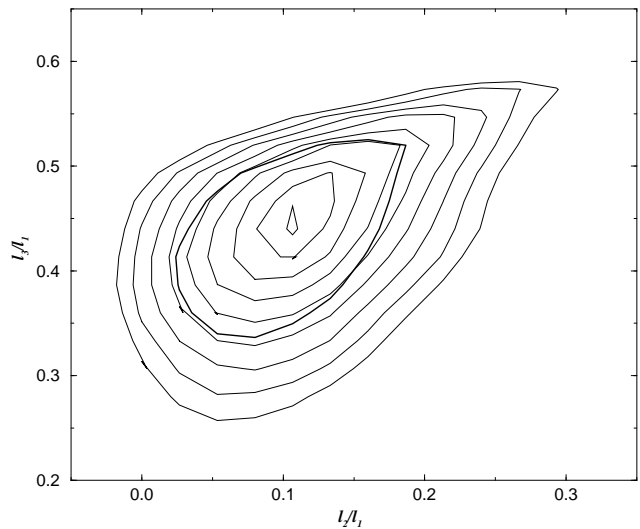, width=7cm, angle=-90}
\end{center}
\caption{\small Contour plot of the scaling function
$f(l_2/l_1,l_3/l_1)$ for the Hecke triangle with $n=7$. Thick line is the 
solution of equation $f(x_2,x_3)=(x_2+x_3-1)/2$.}
\label{f7}
\end{figure}

\begin{figure}
\begin{center}
\epsfig{file=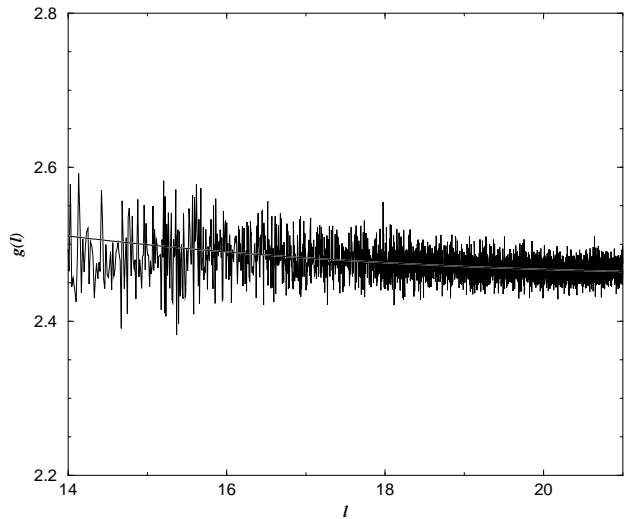, width=7cm, angle=-90}
\end{center}
\caption{\small Mean multiplicity for the Hecke triangle  with $n=7$.
  Solid line is the fit (\ref{fit7}), (\ref{abc}). }
\label{gl7}
\end{figure}

At Fig.~\ref{f7} we present the contour plot of the scaling function
$f(x_2,x_3)$ for the Hecke triangle group with $n=7$ computed from $10^6$ 
points near $l_1=25$. The contour lines correspond to the
sections of the scaling function (normalized so that at the maximum it equals
zero) at heights $-2\cdot 10^{-4}k(2k-1)$ for $k=1,\ldots ,9$.
Numerically from this figure one gets that for the Hecke group with $n=7$ 
the solution of equation $f(x_2,x_3)=(x_2+x_3-1)/2$ do exist and the point 
with the maximum $x_2+x_3$ corresponds approximately to the fourth contour line. 
It means that the density of 
maximal number of different periodic orbit lengths increases as  
\begin{equation}
\rho_{\mbox{\scriptsize diff.
    lengths}}^{(\mbox{\scriptsize maximum})}(l_1) \sim e^{(1-\lambda_7)l_1}
\end{equation}
where $\lambda_7\approx .006$. Correspondingly, the mean
multiplicity of periodic orbit lengths can be estimated (without a prefactor) 
as 
\begin{equation}
\bar{g}(l_1)> e^{.006 l_1}\;.
\label{barg7}
\end{equation}
Though it is an exponential increase, the exponent is so small that at really
accessible lengths $l_1$ of the order of  $20$ it  practically remains a 
constant and the prefactor dominates. At
Fig.~\ref{gl7} we plot the numerically computed mean multiplicity for the Hecke
triangle group with $n=7$ (averaged over interval of traces equal 10). 
Instead of increasing it shows a slow decrease but the best fit to the data 
in the form 
\begin{equation}
\bar{g}=a\frac{e^{bl}}{l^c}
\label{fit7}
\end{equation}
gives 
\begin{equation}
a\approx 3.55\;,\;\;b\approx .007\;,\;\;c\approx .168\;
\label{abc}
\end{equation}
which is larger than (\ref{barg7}). Unfortunately, the limited
interval of lengths and very slow increase of the multiplicity do not 
permit to obtain clear  conclusions.

\begin{figure}
\begin{center}
\epsfig{file=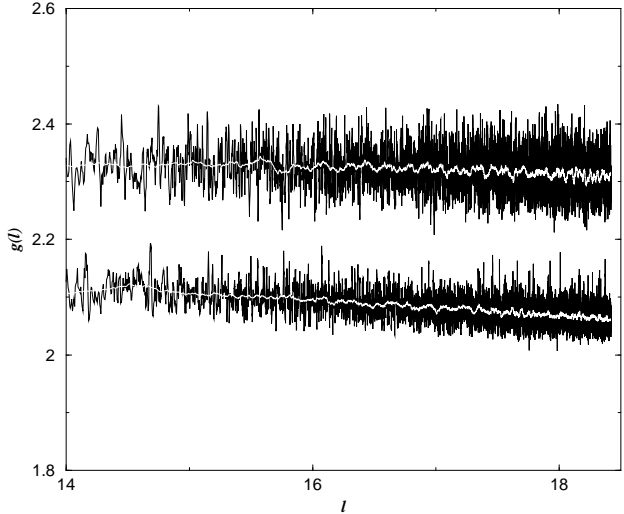, width=7cm, angle=-90}
\end{center}
\caption{\small Mean multiplicities for the Hecke triangles with $n=9$ (the
  upper curve) and $n=11$ (the lower curve). White
  lines represent additional smoothing of the curves.}
\label{gl911}
\end{figure}

At Fig.~\ref{gl911} numerically computed length multiplicities for the Hecke
triangles with $n=9$ and $n=11$ are presented. Similarly to the $n=7$ case
the data indicate a slow decrease which is more pronounced for the $n=11$
triangle. Is this decrease just a lower-length phenomenon or do multiplicities
in these cases tend to a constant cannot be answered from the accessible data.  

We stress that though the data for the Hecke triangles with $n=7$, $n=9$,
and $n=11$ do not show clear increase of mean multiplicities they fluctuate
around values bigger than $2$ which  differs from the usual 
expectation. Also in all figures we present multiplicities averaged over some
length. The true multiplicities fluctuate wildly around the mean confirming
unusual character of the Hecke triangle groups.   

\section{Spectral statistics of non-arithmetic Hecke triangles}\label{statistics}

It is well accepted  that length degeneracy of periodic orbits has a profound
effect on spectral statistics. According to semiclassical theory of spectral
statistics \cite{Berry}, \cite{Berry2} the two-point correlation form factor
for chaotic billiards in the diagonal approximation is 
\begin{equation}
K^{(\mbox{\scriptsize diag.})}(t)=\bar{g}(l(t))t
\label{glt}
\end{equation}
where $\bar{g}(l)$ is the mean multiplicity of periodic orbits with the
length $l$ and $l(t)=4\pi k t$.

For systems without (resp. with) time-reversal invariance $\bar{g}=1$ (resp.
$\bar{g}=2$) and (\ref{glt}) gives the first term of the expansion of the 
two-point correlation form factors for standard random matrix ensembles 
(see e.g. \cite{Bohigas}). 

For models considered in the preceding Sections the mean multiplicity 
$\bar{g}(l)$ increases exponentially as in (\ref{corr}) and the form factor 
calculated in the diagonal approximation differs from the random matrix predictions. 

\begin{figure}
\begin{center}
\epsfig{file=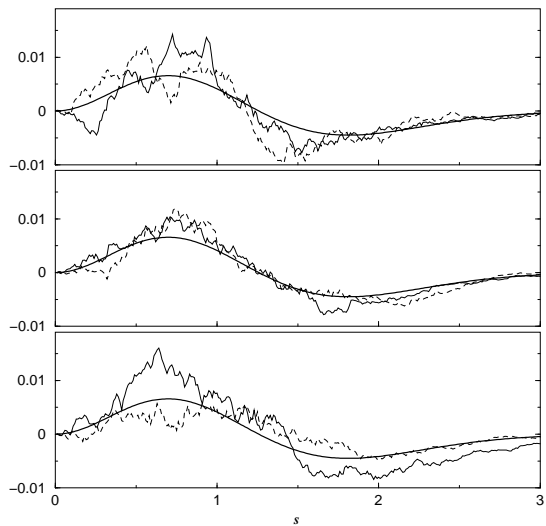, width=7cm, angle=-90}
\end{center}
\caption{\small Differences between integrated nearest-neighbor
  distributions for the Hecke triangles and the Wigner ansatz for
  this quantity (thick solid lines).  Top - the first 10000 levels for the 
  triangle  with $n=5$, middle - the first 20000 for the triangle with $n=7$, 
  bottom - the first 10000 levels for the triangle with 
  $n=12$.  Dashded line - the same quantities but for the
  triangles with angles $10\pi/119$, $10\pi/71$, $20\pi/99$,
  respectively. Thick solid lines at each graph are the difference between 
  the true GOE prediction and the Wigner ansatz. }
\label{ns_5_7_12}
\end{figure}

To consider the spectral statistics we compute numerically
eigenvalues  of the Laplace-Beltrami operator with the Dirichlet conditions
on the boundaries of the Hecke triangles for different values of $n$.
At Fig.~\ref{ns_5_7_12}  we present the differences between the
integrated nearest-neighbor distributions and the Wigner ansatz 
for this quantity ($N_{W}(s)=1-e^{\pi s^2/4}$)
for the Hecke triangles  with $n=5,\;7,\;12$. For comparison on these graphs
thick solid lines indicate the difference between the true GOE formula and
the Wigner ansatz. From the figure it is clearly seen that
spectral statistics  for the Hecke triangles is
quite close to the conjectured statistics of the Gaussian Orthogonal
Ensemble (GOE) of random matrices.
To have an estimate of  statistical errors we compute
numerically the same quantities (dashed lines at Fig.~\ref{ns_5_7_12})
for  non-tesselating triangles of the Hecke
type which have two angles $\pi/2$ and $0$ but instead of the angle $\pi/n$ as
for the true Hecke triangle we take a certain angle $\gamma_n$ sufficiently 
close to it. For $n=5,\;7,\;12$ we choose respectively 
\begin{equation}
\gamma_5=\frac{20\pi}{99}\;,\;\gamma_7=\frac{10\pi}{71}\;,\;
\gamma_{12}=\frac{10\pi}{119}\;.    
\label{gamman}
\end{equation}
Notice that the difference between $\gamma_n$ and $\pi/n$ is quite small
\begin{equation}
\left |\gamma_5-\frac{\pi}{5}\right |\approx 6\cdot 10^{-3}\;,\;
\left |\gamma_7-\frac{\pi}{7}\right |\approx 6\cdot 10^{-3}\;,\;
\left |\gamma_{12}-\frac{\pi}{12}\right |\approx 2\cdot 10^{-3}\;.
\end{equation}  
For all cases (except possible small deviations for $n=12$ which has the 
largest multiplicities) the nearest-neighbor distributions for the Hecke 
triangles agree well with the curves for non-tesselating triangles 

These (and others) figures demonstrate that
the spectral statistics of the non-arithmetic Hecke triangles (even with
quite large degeneracies of periodic orbit lengths)  at small distances is
rather well described by standard random matrix ensembles.  

The contradiction between the observed random matrix statistics of the Hecke
triangles and deviations of correlation functions due to large
multiplicities of periodic orbits (cf. (\ref{glt}) was partially resolved in
\cite{BBGS}. In this paper it was demonstrated that the diagonal
approximation can, strictly speaking, be applied only for very small values of
$t<t_1$. If the mean multiplicity increases like 
$\bar{g}(l)\sim e^{\lambda l}$ with a certain constant $\lambda\leq 1/2$ 
from \cite{BBGS} it follows  that the time of applicability of the diagonal
approximation has the following estimate
\begin{equation}
t_1\sim \frac{1}{1-\lambda}\frac{\ln k}{k}\;.
\end{equation}
During this time the form factor increases exponentially but it can 
reach only a value of the order of 
\begin{equation}
K(t_1)\sim k^{-(1-2\lambda)/(1-\lambda)}\;.
\end{equation}
For arithmetic systems $\lambda=1/2$ (see \cite{BBGS}) and the form factor
for the time of applicability of the diagonal approximation becomes of the
order of $1$ which explains the Poisson character of their spectral
statistics. But for all non-arithmetic groups $\lambda$ is less than $1/2$
and the form factor in the diagonal approximation increases only by a
negative power of $k$. Therefore in the semiclassical limit $k\to \infty$
there is no apparent contradiction between observed $GOE$-type local statistics
and the change of correlation functions due to large multiplicities of periodic
orbits. 

\begin{figure}
\begin{center}
\epsfig{file=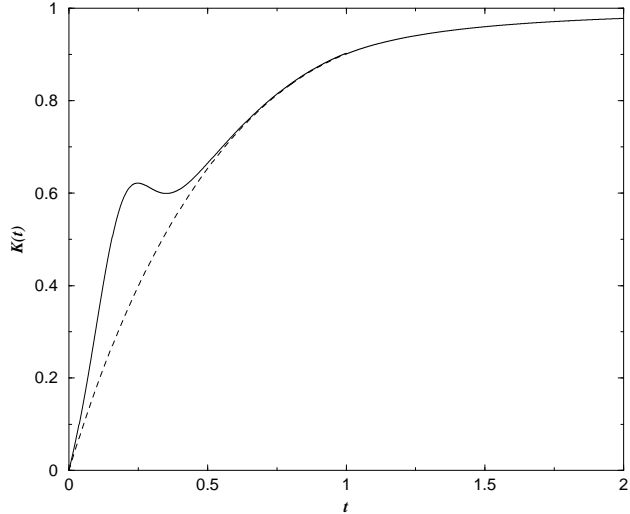, width=7cm, angle=-90}
\end{center}
\caption{\small Schematic form of the two-point correlation form factor for
  non-arithmetic Hecke triangles. Dashed line is the continuation of the GOE
  form factor to small values of $t$.}
\label{kt}
\end{figure}

These arguments suggest that two-point form factors for non-arithmetic
Hecke triangles has the form indicated at Fig.~\ref{kt}. The peak at small
values of $t$ is due to large multiplicities of periodic orbits. The
magnitude of this peak and its position seem to decrease for large $k$. 

\begin{figure}
\begin{center}
\epsfig{file=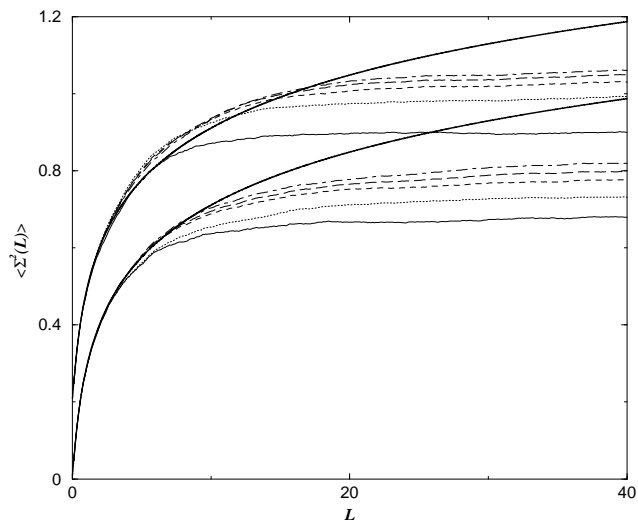, width=7cm, angle=-90}
\end{center}
\caption{\small Smoothed number variance for the Hecke triangle with $n=5$
  (top graphs). Different type of lines corresponds to the number variance
  computed from the first $2000\cdot k$ levels.
  Solid line - $k=1$, dotted line - $k=2$, dashed line - $k=3$,
  long dashed line  - $k=4$, dot-dashed line - $k=5$. Thick line is the 
  GOE prediction.
  Bottom graphs represent the same quantities for the non-tesselating triangle
  with angle $\gamma_5=20\pi/99$. 
  For clarity top graphs  are shifted up by .2 unit.}
\label{sig_5}
\end{figure}

\begin{figure}
\begin{center}
\epsfig{file=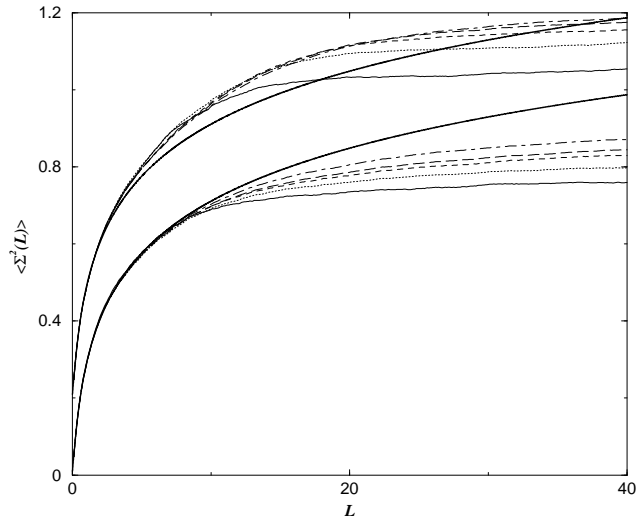, width=7cm, angle=-90}
\end{center}
\caption{\small The same as at Fig.~\ref{sig_5} but for the Hecke triable
  with $n=7$ (top graphs). Different types of lines correspond to the
  averaged number variance computed from the first $4000\cdot k$ levels. 
  Bottom graphs are calculated for the non-tesselating triangle with angle 
  $\gamma_7=10\pi/71$.} 
\label{sig_7}
\end{figure}

\begin{figure}
\begin{center}
\epsfig{file=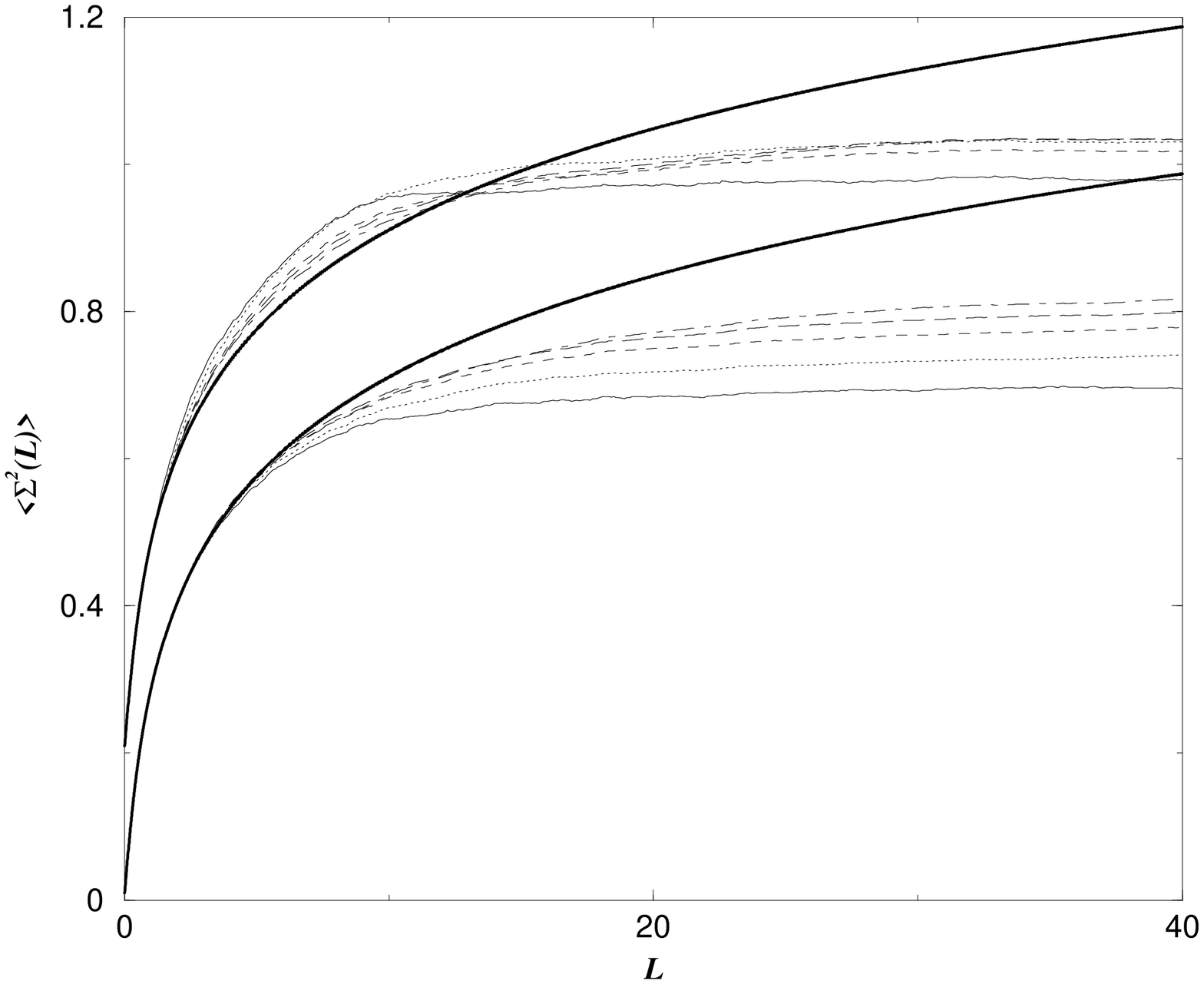, width=7cm, angle=-90}
\end{center}
\caption{\small The same as at Fig.~\ref{sig_5} but for the Hecke triangle
  with $n=12$ (top graphs). Bottom graphs are calculated for the
  non-tesselating triangle with angle $\gamma_{12}=10\pi/119$.} 
\label{sig_12}
\end{figure}

Though deviations from standard statistics should be small when 
$k\to \infty$ the peak indicated at Fig.~\ref{kt} may influence the large
distance spectral properties like the number variance (see e.g.
\cite{Bohigas}). At Fig.~\ref{sig_5}-\ref{sig_12} we present the number
variance  for the Hecke triangles with $n=5,\;7,\;12$
together with the corresponding values for non-tesselating triangles
(\ref{gamman}). Due to large statistical errors in the computation of the
number variance $\Sigma^2(L)$ we found convenient to plot at the figures 
not $\Sigma^2(L)$ itself but its averaged value defined in the following way
\begin{equation}
<\Sigma^2(L)>\equiv \frac{1}{L}\int_0^L\Sigma^2(l)dl\;.
\end{equation}
To demonstrate the evolution of the number variance with increasing the
energy at all figures we present pictures for the averaged number variance
with different number of levels. 

For all non-tesselating (generic) triangles the number variance follows 
the the GOE prediction for small values of $L$ and then saturates, as it 
should be for dynamical systems \cite{Berry}, \cite{Berry2}. 
For the Hecke triangles the
number variance at small $L$ also agrees with the GOE formula but then it
becomes bigger than this reference expression and only later it saturates
but at a value different from the one of the corresponding very close-by 
non-tesselating triangle.  

This overshooting looks as a direct confirmation of the conjectured form of
the two-point correlation form factor (cf. Fig.~\ref{kt}) but careful
calculation of this quantity requires a ressumation of, at least certain,
non-diagonal terms and is beyond the scope of this paper.

\section{Summary}\label{summary}

We demonstrate both numerically and analytically that, at least, certain
non-arithmetic Hecke triangle groups have exponentially large multiplicities of
periodic orbit lengths. 

In groups under consideration matrix elements of group matrices are integers
of an algebraic field of a finite degree and each group matrix gives rise
naturally to $q$ different lengths corresponding to $q$ different
isomorphisms of the basis field. 
The main ingredient of our approach to the problem of periodic orbit length
multiplicity is the investigation of the joint distribution of periodic
orbits with all $q$ transformed lengths fixed. 

We conjecture that this distribution has a scaling form
(\ref{scalingGeneral}) and find the scaling exponent
numerically. For Hecke groups (\ref{two}) with only one non-trivial isomorphism
the general inequality (\ref{inequality}) is sufficient to
demonstrate an exponential increase of the multiplicities. 
Multiplicities obtained by this method are in a good agreement with 
direct numerical calculations.

For general Hecke triangle groups we are not aware of analytical conditions
of the existence of necessary solutions. In all investigated cases (except
for the groups (\ref{two}) with only one non-trivial isomorphism) the
increase of multiplicities numerically is too small to be
observed from direct calculations of periodic orbits but the data fluctuate
around a value bigger than $2$.

The spectral statistics of non-arithmetic Hecke triangles agrees with the
GOE statistics at small distances but deviates from usual expectations
at large distances.

\vspace{1ex}

{\bf Acknowledgments}

\vspace{1ex}

The authors are very  grateful to T. Schmidt for pointing out
Refs.~\cite{Semiarithmetic} and \cite{Paola}, and to O. Bohigas for useful
discussions.

\section*{Appendix A}

To prove the inequality (\ref{inequality}) it is slightly more convenient to
describe conjugacy classes of the Hecke triangle group matrices not by the
code  discussed in Section~\ref{code} but  by a code proposed in 
\cite{Sheingorn} better suitable for analytical calculations. In this code 
the letters for the orientation preserving subgroup of $G_n$ are the following
matrices 
\begin{equation}
g_k(\alpha_n)=U^{k-1}T\;,\;\;\;k=1,\ldots ,n-1\;,
\end{equation}
where $U=TS$ and matrices $T$, $S$ are the translation and
inversion  matrices which generate the whole group $G_n$   
\begin{equation}
T=\left ( \begin{array}{cc} 1&\alpha \\0&1\end{array}\right )\;,\;\;
S=\left ( \begin{array}{cc} 0&-1 \\1&0\end{array}\right )
\end{equation}
with $\alpha=2\cos \pi/n$.

As in the previous code periodic orbits for the Hecke group $G_n$ (with unit
determinant) are free words of letters $g_k$, the only restriction being that 
all cyclic permutations of a word give one orbit.

It is easy to check (e.g. by induction) that
\begin{equation}
g_k=\left ( \begin{array}{cc} a_k&a_{k+1} \\a_{k-1}&a_k\end{array}\right )
\end{equation}
where $a_k\equiv a_k(2\cos \theta)$ are the values of the Chebyshev polynomials
of the second kind 
\begin{equation}
a_k(2\cos \theta)=\frac{\sin (k\theta)}{\sin \theta}\;.
\label{chebychev}
\end{equation}
computed at $\theta=\pi/n$.

Let us introduce the following definition.  
We say that a function $f(x)$ has the H-property  with a separating point 
$h$ if 
\begin{equation} 
 |f(x)|\leq f(h)\; \mbox{ for all }\;|x|\leq h\;.
\end{equation}  
The importance of this notion follows from the fact that
if $f_1(x)$ and $f_2(x)$ both have the H-property
with separating point $h$ then $f_1(x)f_2(x)$ and $f_1(x)+f_2(x)$
also have the H-property with the same separating point. In particular, if
one has a set of matrices whose elements all have the H-property with a
separating point $h$ then all products of these matrices also have the
H-property with the same separating point.

Let us prove first that matrix elements $a_k$ with $k=0,\ldots ,n$ have 
the H-property  with the separating point $h=2\cos(\pi/2n)$. Indeed
\begin{equation}
\left |a_k(2\cos \theta)\right |\leq \frac{1}{\sin \theta}\;.
\end{equation}
When $0\leq \theta\leq \pi/2$ the equality sign in this inequality 
holds at the points 
\begin{equation}
\theta_m=\frac{\pi}{2k} m
\end{equation}
where $m$ is an odd integer $1\leq m\leq k $.

Therefore when $\theta\geq \pi/2k$
\begin{equation}
\left |a_k(2\cos \theta)\right |\leq \frac{1}{\sin \theta}
  \leq a_k(2\cos \frac{\pi}{2k})\;.
\end{equation}
Moreover, $a_k(2\cos \theta)$ is a decreasing function  when $0<\theta<\pi/2k$
and
$a_k(2\cos \theta)\leq a_k(2\cos \pi/2n)$ when $\pi/2n\leq \theta \leq \pi/2k$ 
(and, of course, $k\leq n$). 
Together these two inequalities prove that $a_k$ with $k\leq n$ have the 
H-property with separating point $h=2\cos(\pi/2n)$.  
 
Second, $a_k(2\cos \pi p/ n)=a_{n-k}(2\cos \pi p/n)$ for odd
integer $p$. It means that for all isomorphisms (\ref{iso}) $a_k(x)$ equals
$a_{n-k}(x)$ and only $a_k$ with $k\leq n/2$ are independent. Hence, $a_k$ 
for all isomorphisms of the defining equation can be considered 
as polynomials of degree not greater than
$n/2$ and one can choose for all $a_k(2\cos\pi p/n)$ with  $k\leq n$  the same 
separating constant $h=2\cos(\pi/n)$. 

Third, as was stated above, all matrix elements
obtained by taking the products of arbitrary number of matrices $g_k$ also
have the H-property with the separating constant $h=2\cos(\pi/n)$. 

Combining all these arguments one proves that for all isomorphisms of the
basis field traces of the Hecke group matrices have the H-property with
the same separating constant. Because  
\begin{equation}
\left |\cos\frac{\pi k}{n}\right |<\cos \frac{\pi}{n}
\end{equation}
for all $k\neq 0,1, n$ one gets that modulus of traces of the 
Hecke triangle group
matrices decrease for all non-trivial isomorphisms of the basis field thus
proving the inequality (\ref{inequality}). For matrices with determinant
equal $-1$ the same inequality follows by computing the square of such matrices
because the periodic orbit length for the square of any matrix is twice
the length corresponding to the initial matrix. 

After this paper has been completed we become aware of
Ref.~\cite{Semiarithmetic} where the inequality (\ref{inequality}) was 
proved for all groups which permit the so-called modular embedding. From 
Ref.~\cite{Paola} it follows that all triangle discrete groups belong to 
this class. Therefore, the inequality  (\ref{inequality}) is valid for all 
triangle groups  (and not only for the Hecke triangle groups considered in this 
paper).   

\section*{Appendix B}

The purpose of this Appendix is to give arguments in favor of the
representation (\ref{scaling}) of the joint probability density of periodic 
orbits with all transformed lengths fixed. 

For discrete groups periodic orbits can be obtained from product of certain
number of matrices. Let us consider in  a given code the product of  $n$  
basis matrices
\begin{equation}
A(n)=A_n\cdot A_{n-1}\cdots  A_1\;.
\label{an}
\end{equation}
The total number of matrices with $n$ symbols for a general code is exponential 
\begin{equation}
\rho (n)\stackrel{n\to \infty}{\longrightarrow} \frac{e^{hn}}{n}
\end{equation}
where $h$ is a constant called the topological entropy.

The length of periodic orbit is related with matrix $A$ asymptotically 
as 
\begin{equation}
l=2\ln \mbox{ Tr }A \;.
\end{equation}
Therefore, matrices representing periodic orbits can be considered as the
result of a random process where matrices $A_k$ are chosen randomly from a
code grammar. The probability distribution of lengths for products of $n$ 
such matrices is defined as the ratio of  number of matrices  with lengths 
in the interval $[l,l+dl]$ divided by total number of matrices.
    
This distribution under quite general conditions \cite{Furstenberg}, 
\cite{Goldsheid} has  the Gaussian form 
\begin{equation}
P_n(l)=\frac{1}{\sqrt{2\pi}\sigma_n}e^{-\frac{(l-l_n)^2}{2\sigma_n^2}}
\end{equation}
where  
\begin{equation}
l_n=\lambda_0n\;,\;\;\;\sigma_n^2=\sigma_0^2 n\;.
\end{equation}
One of possible applications of such distribution is the calculation of 
the total density of periodic orbits of length $l$ (see e.g. \cite{Sieber})
\begin{equation}
\rho(l)=\int \frac{e^{hn}}{n}P_n(l)dn\;.
\label{rhol}
\end{equation}
When $l\to \infty$ the integral can be computed in the saddle point 
approximation and the total density of periodic orbits has exponential
asymptotics
\begin{equation}
\rho(l)=\frac{e^{\kappa l}}{l}
\end{equation}
where 
\begin{equation}
\kappa=\frac{\lambda_0-\sqrt{\lambda_0^2-2h\sigma_0^2}}{\sigma_0^2}\;.
\end{equation}
For groups considered in the paper all matrix elements belong to
an algebraic field of a finite degree which has $q$ different isomorphisms.
It means, in particular, that each product of $n$ group matrices
$A(n)$ as in (\ref{an}) gives rise to $q$ different lengths $l_i(n)$ obtained
by applying each isomorphism $\varphi_i$ to $A(n)$
\begin{equation}
l_i(n)=2\ln |\mbox{Tr}\varphi_i(A(n))|\;.
\end{equation}
On the other hand $\varphi_i(A(n))$ can be obtained as the product of $n$
transformed matrices $\varphi_i(A_k)$ as in (\ref{an}). Therefore according
to the above theorem each variable $l_i$ is a random variable whose
distribution also has  asymptotically the Gaussian form 
\begin{equation}
P_n(l_i)=\frac{1}{\sqrt{2\pi n}\sigma_i}
               e^{-\frac{(l_i-\lambda_i n)^2}{2n\sigma_i^2}}
\end{equation}
with certain parameters $\lambda_i$ and $\sigma_i$ having the  meaning of the
mean value and the variance of $l_i(n)$.

Let us conjectured that the mutual distribution of all $l_i(n)$ together is 
also Gaussian 
\begin{equation}
P_n(\vec{l}\ )=\frac{\sqrt{\det M}}{(2\pi )^{q/2}}
\exp \left (-\frac{1}{2n}\sum_{ij=1}^q M_{ij}
(l_i-\lambda_i n)(l_j-\lambda_jn)\right )
\end{equation}
with a certain positive definite matrix $M_{ij}$.

Analogously to Eq.~(\ref{rhol}) the total density of orbits with fixed $l_i$ is
\begin{equation}
\rho(\vec{l}\ )=\int_1^{\infty} \frac{e^{hn}}{n}P_n(\vec{l}\ )dn\;.
\end{equation}
As above this integral can be computed in the saddle point approximation
valid at large $\vec{l}$ and the result is
\begin{equation}
\rho(\vec{l}\ )=\frac{\sqrt{\det M}}{(2\pi )^{(q-1)/2}(A(C-2h))^{1/4}}\exp
\left (B-\sqrt{A(C-2h)}\right )
\label{saddle}  
\end{equation}
where
\begin{equation}
A=\sum_{i,j=1}^qM_{ij}l_il_j\;,\;\;B=\sum_{i,j=1}^qM_{ij}l_i\lambda_j\;,\;\;
C=\sum_{i,j=1}^qM_{ij}\lambda_i\lambda_j\;.
\end{equation}
The exponent in (\ref{saddle}) is an homogeneous function of $l_j$ and after 
the rescaling $x_j=l_j/l_1$ one obtains the scaling ansatz 
(\ref{scalingGeneral}) with a specific scaling function which leads to 
exponential asymptotics of the joint probability distribution as it seems 
suggested by numerics (cf. Figs.~\ref{f5}-\ref{f12}).

The main drawback of such approach is that the theorem about the Gaussian
form of the distribution of the product of $n$ random matrices is valid only
near the maximum of the distribution.  But the term $e^{hn}$
in (\ref{rhol}) and (\ref{saddle}) shifts the saddle point far from the
maximum and there exist no general arguments which would imply the
smallness of  corrections to the parabolic form of the exponent. For certain
groups and special codes it seems that one can ignore such corrections in a
region of interest  but, in general, corrections are large and one has to
rely on the numerics as it was done in the main part of the paper.    

\end{document}